# Electromagnetic Proton Form Factors

M. Y. Hussein
Department of Physics
College of Science
University of Bahrain

**Abstract**
The electromagnetic form factors are crucial to our understanding of the proton's internal structure, and thus provide a strong constraint of the distributions of charge and magnetization current within the proton. We adopted the quark-parton model for calculating and understanding the charge structure of the proton in terms of the electromagnetic form factors. A remarkable agreement with the available experimental evidence is found.

The understanding of nucleon structure is one of the central problems of hadronic physics [1]. The complex structure of hadrons can be described in a convenient way in terms of form factors. The form factors of the proton and neutron are considering as fundamental properties of the nucleon, and gave a critical test ground for models based on QCD. A detailed knowledge of these form factors is essential to our understanding of the electromagnetic response functions of nuclei. Measurements of the electric and magnetic form factors, $F_E$ and $F_G$, as function of the four-momentum transfer $q^2$ provide experimental information relating to the nucleon structure.

The (elastic or inelastic) electron-hadron scattering is the traditional way to determine the form factors, and it allows a direct comparison with the theory. The experimental determination of the proton form factor has been carried out by directing high energy electron beams at a hydrogen target, and making precision measurements of the momentum and angle of scattered electrons by means of magnetic spectrometers. The first experiments demonstrating the deviation of the scattering from the expected for point particles and thus measuring the form factors were carried out by Hofstadter and his calibrators in 1961, at Stanford, and have since been extended in numerous laboratories [1]. Recent measurements at the Jefferson Laboratory essentially improved the experimental data concerning the elastic form factors of protons and deuterons up to relatively large values of momentum transfer [2,3,4]. A large body of precise data obtained primarily from elastic scattering of electrons by protons and deuterons now exists for these form factors. Recent data for large space-like momentum transfer are in remarkable agreement with the prediction for $q^2$ as small as $5(GeV)^2$ [5]. The study of the internal structure of light hadrons can be carried on with a high intensity electron machine and a large program is under way to measure the form factors of the nucleon resonance [6,7,8]

Nucleon's charge structure has been a subject of extensive investigation for the past years [9]. There have been numerous attempts aiming at a plausible explanation of the charge structure and form factor using a variety of techniques including, for instance, the empirical charge distributions [10] on one hand and the quark-parton models on the other [11], which besides being the easiest to understand has had tremendous success to its credit.

The present investigation is an attempt to understand and account the electric form factors for the proton within the general prescriptions of quark-parton model. Our consideration is based on the hypothesis that the proton has a disc-like appearance due to the influence of Lorentz contraction when the energy of the incident beams is in the high-energy region. In view of this geometrical picture with finite but small disc thickness, it is essential to consider not only the transverse distribution functions, $h_i(x,b)$, but also a longitudinal spatial distribution, $g_i(y,z)$, with a suitable cut off $z_0 = 0.017\,fm$. The complete quark parton distribution function can be written in a separable form as:

$$\rho_i(x,b;y,z) = h_i(x,b)g_i(y,z) \qquad (1)$$

With $h_i(x,b) = [\pi A(x)]^{-1} f_i(x) e^{-[b^2/A(x)]}$, $g_i(y,z) = [z_0(1-\frac{z_0^2}{3B(y)})]^{-1} f_i(y)(1-\frac{z^2}{B(y)})$ are properly normalized spatial distribution functions, $A(x) = e^{-\beta x}$, $f_i(x)$ is the mean density of quark $i$ with momentum fraction $x$ and $f(y)$ is the mean density of quark $i$ with momentum friction $y$ such that $x^2 + y^2 = 1$.

The charge form factor $F_p(q^2)$ for the proton can be obtained by taking the Fourier transform of the quark-parton distribution function $\rho_i(x,b;y,z)$ as:

$$F_p(q^2) = \int dx \int dy \delta(x^2 + y^2 - 1) \sum_i e_i F_i(q^2, x, y) \qquad (2)$$

Where $F_i(q^2, x, y)$ is given by:

$$F_i(q^2, x, y) = \int d^3 r \rho_i(x,b;y,z) e^{iq.r} \qquad (3)$$

$$F_i(q^2, x, y) = H_i(q^2, x, y) R_i(q^2, x, y) \qquad (4)$$

With

$$H_i(q^2, x, y) = \int dz g_i(y,z) e^{iqxz} \qquad (5)$$

$$R_i(q^2, x, y) = \int d^2 b h_i(x,b) e^{iqyb} \qquad (6)$$

Making use of the definition of distribution function $g_i(y,z)$ in equation (5) and with proper integration over $dz$, we get:

$$H_i(q^2, x, y) = [z_0(1 - \frac{z_0}{3B(y)})]^{-1} f_i(y) e^{\frac{iqxz_0}{2}} [P + iS] \qquad (7)$$

Now making use of the definition of distribution function $h_i(x,b)$ in equation (6) and with proper integration over $d^2b$ we get:

$$R_i(q^2, x, y) = f_i(x) e^{-\frac{1}{4}q^2 y^2 A(x)} \qquad (8)$$

These values of the integrals $H_i(q,x,y)$ and $R_i(q,x,y)$ lead to the final result of the form factor:

$$F_p(q^2) = \int dx \int dy \, \delta(x^2 + y^2 - 1) \sum_i e_i f_i(x) f_i(y) [z_0(1 - \frac{z_0}{3B(y)})]^{-1} f_i(y) e^{\frac{iqxz_0}{2}} [P + iS]$$
$$e^{-\frac{1}{4}q^2 y^2 A(x)} \qquad (9)$$

Integrating equation (9) over $dy$ using the property of the Dirac $\delta$-function we get:

$$F_p(q^2) = \int_0^1 dx \frac{B(\sqrt{1-x^2})}{z_0(3B(\sqrt{1-x^2}) - z_0^2)} \frac{3Q_p(x)}{2\sqrt{1-x^2}} [P^2 + S^2]^{\frac{1}{2}} e^{-\frac{1}{4}q^2(1-x^2)A(x)} \qquad (10)$$

Where $Q_p(x)$ represents the compound quark density function for the proton and which has the form:

$$Q_p(x) = \frac{2}{3} u_v(x) u_v(y) - \frac{1}{3} d_v(x) d_v(y) \qquad (11)$$

Calculations are performed for the proton form factors as a function of the momentum transfer $q^2$ using equation (10). In our calculations, we use the most recent structure functions from MRST [12]. The value of $\beta$ is determined from the experimentally known value of $\langle r_p^2 \rangle$ which is taken to be $0.707 \, fm^2$ [13]. A comparison with the corresponding experimental values of $F_p(q^2)$ is extremely remarkable. Our results are particularly interesting in that they are the closest to the observed data in comparison to the performance to other related quark parton distribution functions.

In Ref. [14], the value of the proton form factor $F_p(q^2)$ has been parameterized in dipole form:

$$F_p(q^2) = \frac{1}{(1 + q^2 / 0.71 GeV^2)^2} \qquad (12)$$

And it is not rigorously identified as magnetic or electric, proton or neutron. From quark counting rules considerations, the dipole form of the nucleons form factors has been taken until recently as universal, consistent with the experimental data the nucleon form factors.

Before the JLAB measurement [15], the experimental data about electron-nucleon scattering based on Resenbluth separation were consistent with the representation. The data from [15] can be fitted by:

$$F_p(q^2) = \frac{1}{(1+q^2/0.71 GeV^2)^2} \frac{1}{(1+q^2/4.8 GeV^2)} \qquad (13)$$

Where the second factor explicitly shows the derivation from the dipole form.

In Figure 1 we illustrate the behavior of the proton form factors $F_p(q^2)$ as a function of momentum transfer $q^2$.

In conclusion, the results are particularly interesting in that they are the closest to the observed data in comparison to the performance of other related QPDF's. For higher values of $q^2$, the experimental data for scattering are still not available and hence the model predictions can't be checked in detail. There are, however, a few points that needed be looked into prior to a meaningful description of the results. The experimental information demands a critical scrutiny in view of the uncertainties arising from the difficulties associated with the choice of the target. Also we have used a special form of functions for these choices are none other than simplicity and for effectively reducing the number of free parameters. It is quite conceivable that other functional forms with a proper choice of the variables will make the results much better.

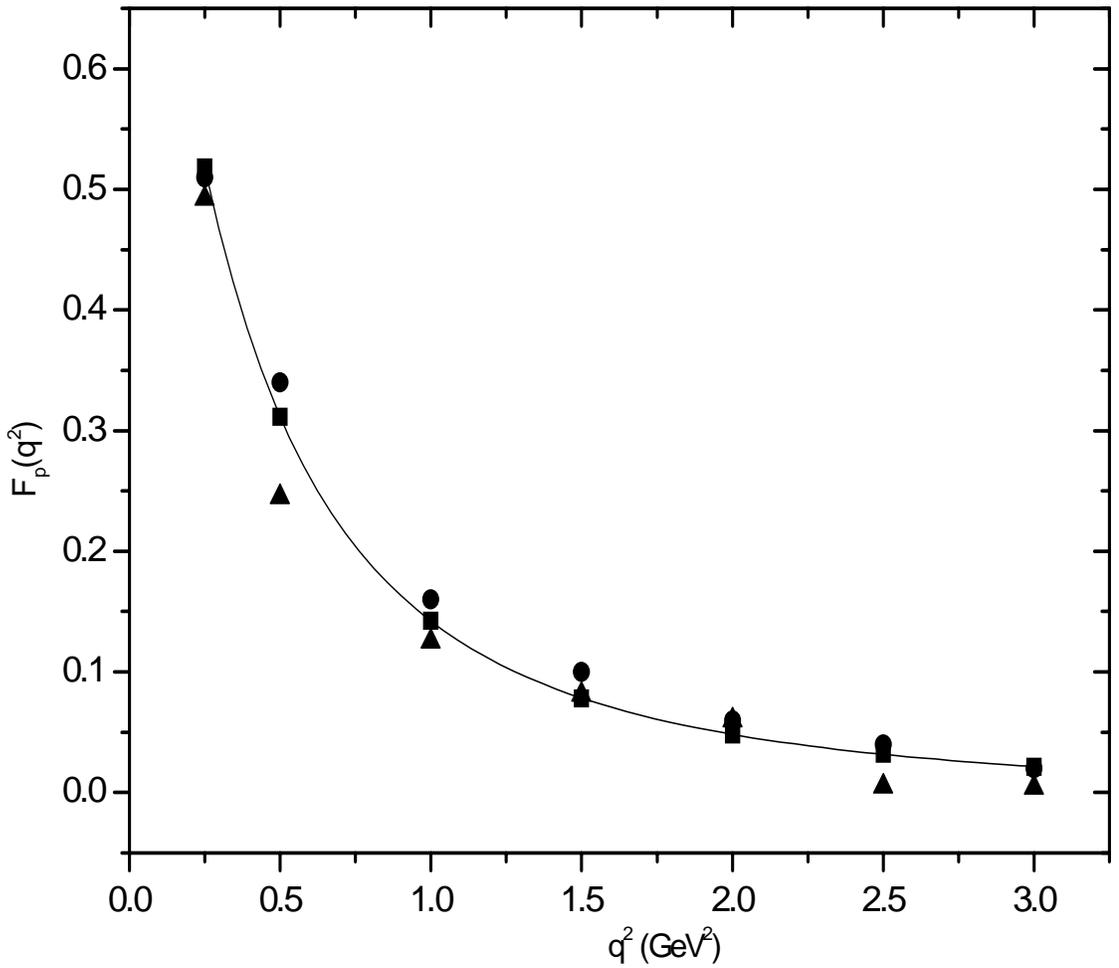

Figure 1: Proton electric form factors as function of the momentum transfer $q^2$. The electric form factors from Ref. [2] (solid squares). The solid circles is the parametrization for $F_p(q^2)$ form Eq. (13). The solid triangles is the theoretical calculations of the proton form factors from Eq. (10) using MRST quark distribution functions.